\documentclass[english,aps,twocolumn,prl]{revtex4}
\usepackage[T1]{fontenc}
\usepackage[latin9]{inputenc}
\usepackage{graphicx}

\makeatletter
\@ifundefined{textcolor}{}
{%
 \definecolor{BLACK}{gray}{0}
 \definecolor{WHITE}{gray}{1}
 \definecolor{RED}{rgb}{1,0,0}
 \definecolor{GREEN}{rgb}{0,1,0}
 \definecolor{BLUE}{rgb}{0,0,1}
 \definecolor{CYAN}{cmyk}{1,0,0,0}
 \definecolor{MAGENTA}{cmyk}{0,1,0,0}
 \definecolor{YELLOW}{cmyk}{0,0,1,0}
}

\makeatother

\usepackage{babel}
\begin{document}

\preprint{\pagebreak{}This line only printed with preprint option}

\title{Qualitative breakdown of the unrestricted Hartree-Fock energy}

\author{Paula Mori-Sánchez\footnote[4]{paula.mori@uam.es}}

\affiliation{Departamento de Química and Instituto de Física de la Materia Condensada
(IFIMAC), Universidad Autónoma de Madrid, 28049, Madrid, Spain}

\author{Aron J. Cohen\footnote[1]{ajc54@cam.ac.uk}}

\affiliation{Department of Chemistry, Lensfield Rd, University of Cambridge, Cambridge,
CB2 1EW, UK }
\begin{abstract}
The stretching of closed-shell molecules is a qualitative problem
for restricted Hartree-Fock that is usually circumvented by the use
of unrestricted Hartree-Fock (UHF). UHF is well known to break the
spin symmetry at the Coulson-Fischer point, leading to a discontinuous
derivative in the potential energy surface and incorrect spin density.
However, this is generally not considered as a major drawback. In
this work, we present a set of two electron molecules which magnify
the problem of symmetry breaking and lead to drastically incorrect
potential energy surfaces with UHF. These molecules also fail with
unrestricted density-functional calculations where a functional such
as B3LYP gives both symmetry breaking and an unphysically low energy
due to the delocalization error. The implications for density functional
theory are also discussed.
\end{abstract}
\maketitle

\section{Introduction}

The restricted Hartree-Fock method (RHF) is one of the cornerstones
of quantum chemistry for the description of closed-shell molecules.
However, RHF breaks down when a bond in a closed shell system is stretched.
One way around this problem, that works in many cases, is to use the
unrestricted Hartree-Fock method (UHF), which allows different spatial
orbitals for the $\alpha$ and $\beta$ electrons and avoids the error
of RHF by breaking the spin symmetry. For the prototypical example
of the H$_{2}$ molecule, UHF gives the same description as RHF around
equilibrium, and as the bond is stretched well beyond the minimum,
it goes through the Coulson-Fischer point \cite{Coulson49386} at
which the spin symmetry breaks to dissociate into an unrestricted
solution with the correct energy and one $\alpha$ electron on one
hydrogen and one $\beta$ electron on the other hydrogen. 

Due to the breaking of the spin symmetry there is a discontinuous
derivative of the energy with respect to geometry at the Coulson-Fischer
point. Furthermore, the UHF broken symmetry solution is not an eigenstate
of the total spin operator and can give not just spin densities, but
total densities, that do not obey the overall symmetry of the Hamiltonian.
This has been observed in many cases; from non-spherical atomic densities
for open shell $p$ atoms \cite{Savin96327,Cohen08121104}, symmetry
breaking in the uniform electron gas and related systems \cite{Mendl14125106,Loos09062517,Thompson04201302}
and stretched He$_{2}^{+}$ \cite{Mori-Sanchez1414378}. However,
for the purposes of this paper, we will assume that none of these
is a major issue. In H$_{2}$, the Coulson-Fischer point is well after
the equilibrium geometry, but in molecules such as F$_{2}$ \cite{Gordon871,Purwanto08114309}
and O$_{2}^{2+}$ \cite{Nobes1991216}, it can occur at much shorter
bond lengths and even before the RHF minimum. Symmetry breaking also
plays an important role in the calculation of antiferromagnetic exchange
couplings that are important in transition metals complexes \cite{Noodleman1986131}.
In this case, a simple model often used, which also shows a symmetry
breaking early in its binding curve, is the stretching of HHeH\cite{Caballol977860}.

\begin{figure}[b]
\includegraphics[width=1\columnwidth]{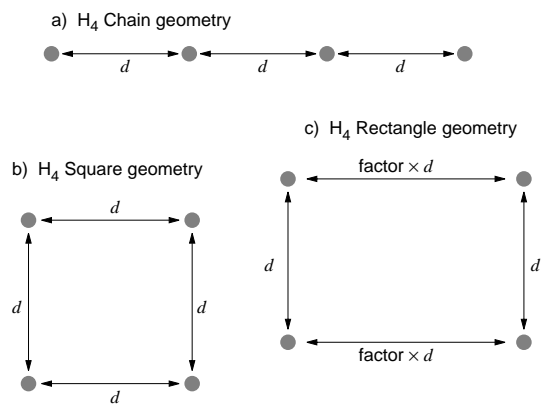}

\caption{The geometries of H$_{4}$/H$_{4}^{2+}$ studied in this paper. Several
different rectangles are considered, each with a different factor.
Increasing $d$ leads to a uniform stretching of all of these molecules.\label{fig:The-geometries}}
\end{figure}

\begin{figure*}
\includegraphics[angle=-90,width=1\textwidth]{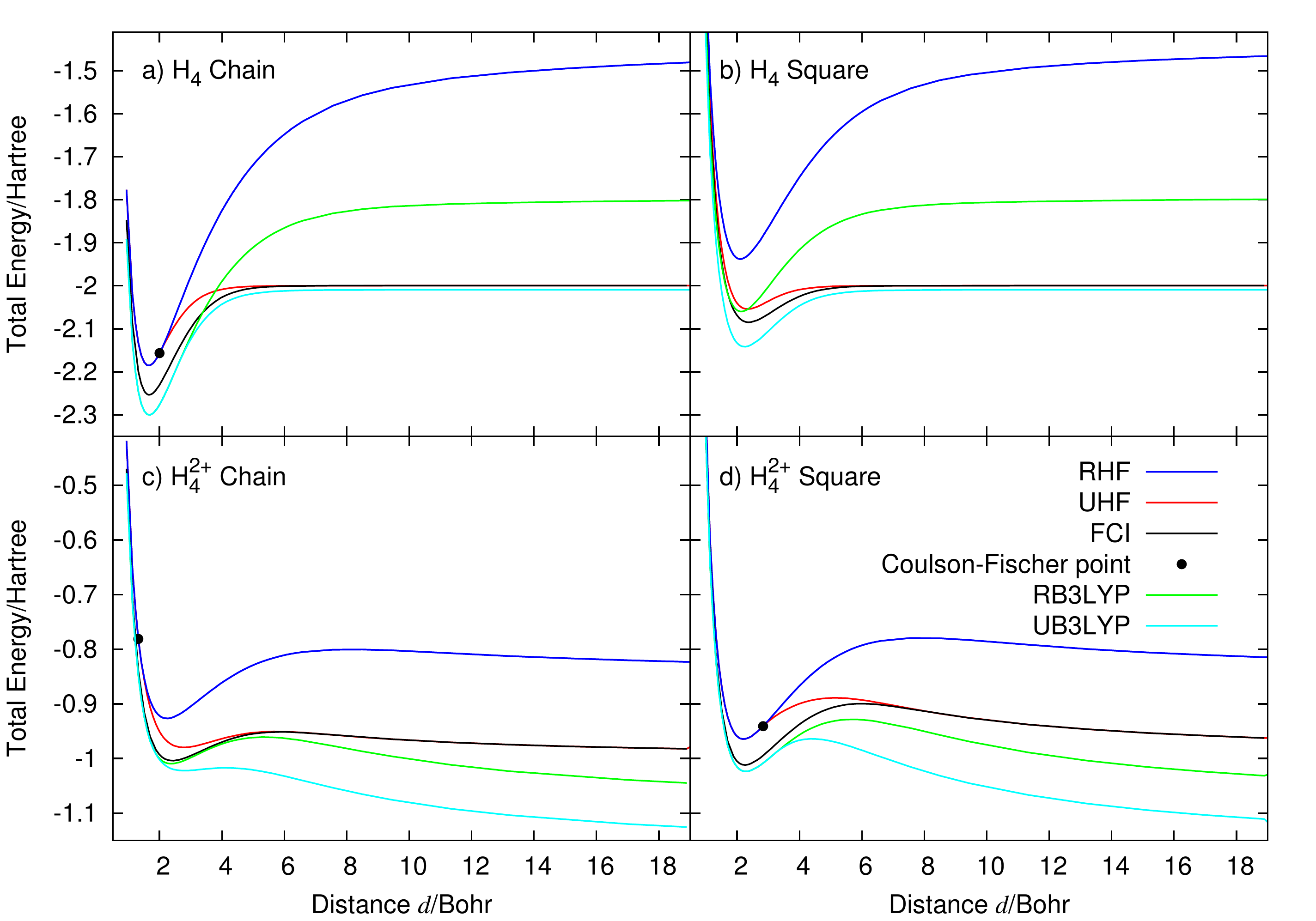}

\caption{The uniform stretching of H$_{4}$ and H$_{4}^{2+}$ in the chain
and square geometries calculated with RHF, UHF, RB3LYP, UB3LYP and
FCI.}
\end{figure*}

\begin{figure*}
\includegraphics[angle=-90,width=1\textwidth]{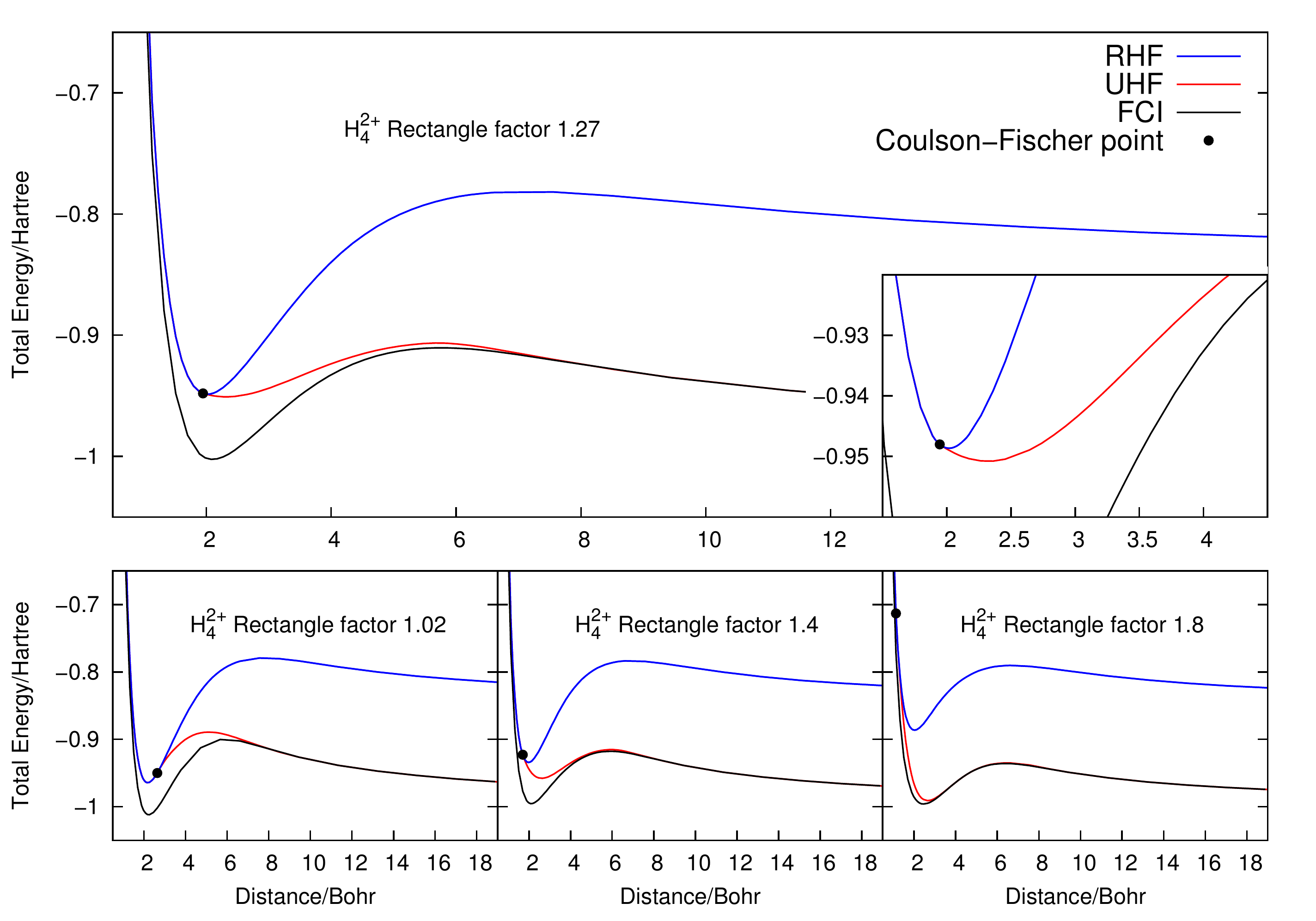} 

\caption{Stretching of a set of rectangular geometries of H$_{4}^{2+}$. Factor
is the ratio between the two sides, and is constant as the distance
is varied.\label{fig:Rectangle H42+} The inset shows a magnified
view of the UHF minimum for the rectangle with factor 1.27.}
\end{figure*}

There is still the open question of whether symmetry breaking is a
good idea or it is just avoiding an error of the RHF method in stretching
bonds \cite{Jimenez112667}. This is true for perturbation theory
\cite{Knowles883097,Nobes1991216,Jarzecki984742} but is also a particularly
important question in relation to density functional theory (DFT).
In this paper, we want to show a set of two-electron systems which
reveal a clear qualitative breakdown in the total energy of restricted
Kohn-Sham DFT (RKS), unrestricted Kohn-Sham DFT (UKS) and, remarkably,
also UHF.

\section{Results and Discussion}

Simple molecules of hydrogen atoms as in Fig. \ref{fig:The-geometries}
are studied. All calculations are carried out using Gaussian09 \cite{g09}
and the cc-pVQZ basis set \cite{Dunning891007}. The optimization
of the wavefunction is done using the standard DIIS procedure checking
for wavefunction instabilities \cite{Paldus702268} as well as the
quadratic convergent SCF method\cite{Bacskay81385}. We have also
independently checked all the results by comparing with our own code
based on the direct minimization of the total energy imposing orbital
orthonormality \cite{Goedecker971765,Cohen02409}, to try and ensure
that we have found the best SCF orbitals at each geometry. The full
configuration interaction (FCI) calculations are carried out using
the \texttt{fci} code of Knizia and Chan developed in their work on
density matrix embedding theory \cite{FCIcode}.

First, in Fig. 2a, consider the linear H$_{4}$ chain and its stretching,
with an equal spacing between the hydrogen atoms. The neutral system
with four electrons has very similar binding curves to the well known
H$_{2}$ molecule. Fig. 2a shows the usual behavior of RHF, with a
large static correlation error, and UHF following RHF beyond equilibrium
up to around 2 bohr until a Coulson-Fischer point, where it breaks
the spin symmetry to give the correct dissociation limit compared
to FCI. Consider now a square geometry for H$_{4}$ with four electrons
and look at the uniform stretching (Fig. 2b). The behavior of RHF
and UHF is markedly different. In this, case RHF is massively wrong
at nearly all geometries and consequently UHF breaks the spin symmetry
at all distances and gives good agreement with FCI, although missing
the dynamical correlation, to reach the correct energy at stretching.
Here, although the spin density is wrong, the total density and energy
are acceptable. UB3LYP performs reasonably in both cases (Fig 2a and
2b).

Now, examine the same systems but with two electrons removed, that
is H$_{4}^{2+}$ linear chain and square geometries. Again, quite
different performance is observed. RHF exhibits the usual behavior
in the square geometry, whereas it is completely wrong at all distances
in the chain structure. In these cases, UHF either agrees well with
FCI (H$_{4}^{2+}$ chain, Fig. 2c) or has the usual Coulson-Fischer
symmetry breaking (H$_{4}^{2+}$ square, Fig 2d). Although this Coulson-Fischer
symmetry breaking is not appealing, the potential energy curves are
reasonable. The poor performance of both RB3LYP and UB3LYP for H$_{4}^{2+}$
in these geometries must be noted, which is due to the delocalization
error \cite{Mori-Sanchez08146401}, related to the non-integer number
of electrons on each atom.

\begin{figure*}
\includegraphics[angle=-90,width=1\textwidth]{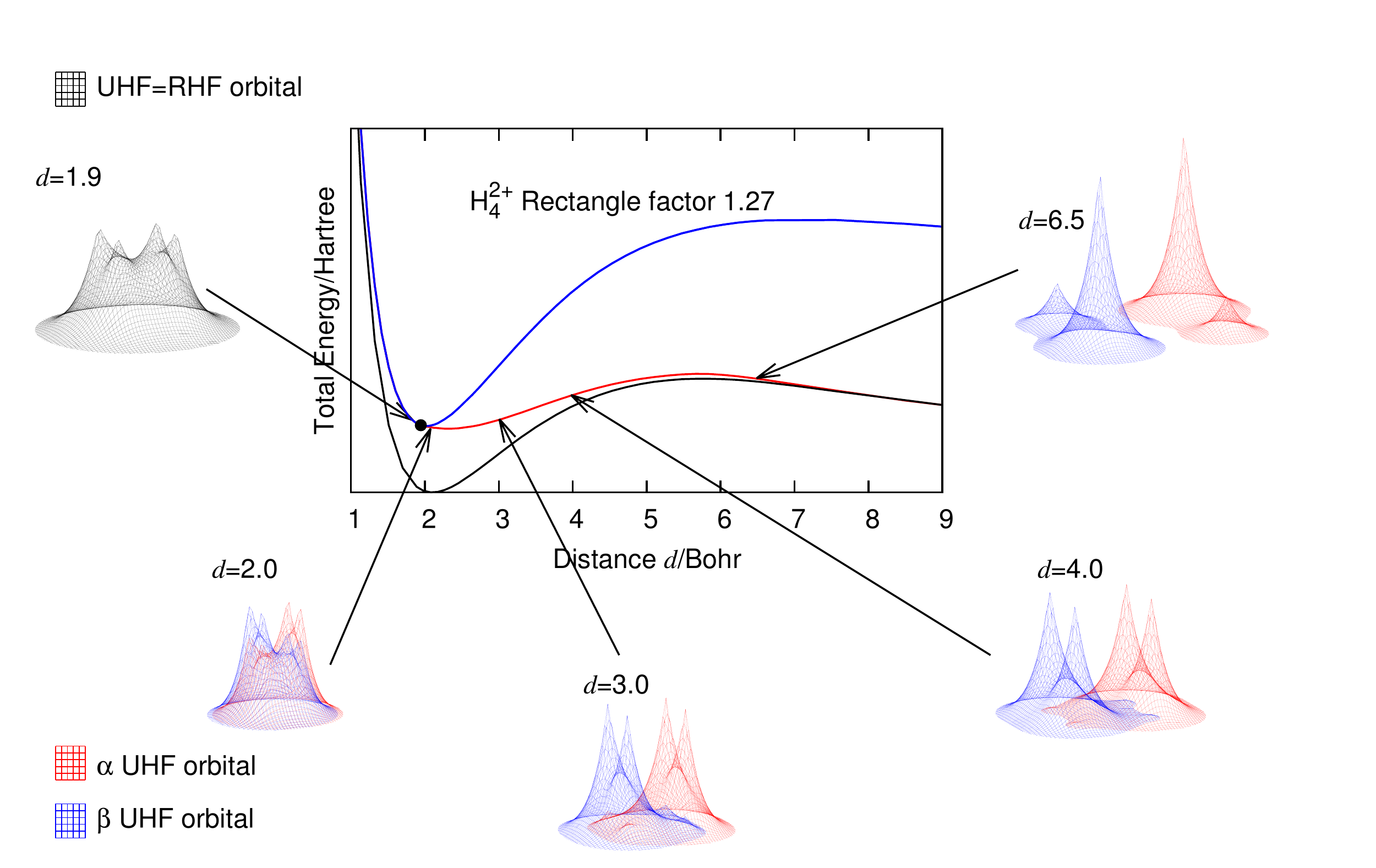}\caption{Plot of the UHF orbitals for the stretching of H$_{4}^{2+}$ rectangular
geometry with ratio between the sides of factor=1.27 shown for different
values of $d$. The red and blue surfaces show the $\alpha$ and $\beta$
orbitals respectively, at $d=1.9$ $\phi_{\alpha}=\phi_{\beta}$ and
this is depicted by a black surface.\label{fig:orbitals}}
\end{figure*}

The behaviors of RHF and UHF seen in the stretching of the chain and
square in Figs. 2c and 2d are somewhat limiting cases (a tetrahedron
is also similar to the square). We can consider ways to go smoothly
in between the two types of behavior if we take H$_{4}^{2+}$ in a
rectangular geometry where the ratio between the two sides is a constant
factor as the system is stretched. A factor of 1.0 corresponds to
the square geometry and a factor of 2.0 leads to the same nearest
neighbor and next nearest neighbor distances as the linear chain.
Fig. \ref{fig:Rectangle H42+} shows a range of factors between 1.0
and 2.0. First, it should be noted the similar nature of the FCI curves
between the plots, which indicates that for real electrons there is
not a great deal of difference between the different geometries. Also
for RHF there is not much difference between the different ratios.
It is only for UHF that the curves depend greatly on the ratio. For
example, for a ratio of 1.27, the symmetry breaking Coulson-Fischer
point is near the minimum of the RHF curve. This leads to a very poor
UHF curve that has fundamental problems. It gives a minimum with a
very large qualitative error, and calculations with smaller basis
sets can even give curves with double minima.  

The nature of the symmetry breaking of the orbitals is shown in Fig.
\ref{fig:orbitals}. At short distances the $\alpha$ and $\beta$
orbitals are equal and both over all the atoms, then at the Coulson
Fischer point the symmetry breaks along the longer distance such that
the $\alpha$ electron is on the two right atoms and the $\beta$
electron is on the two left atoms, and finally at longer distance
a further symmetry breaking occurs along the shorter distance so that
in the limit of large $d$ UHF gives the $\alpha$ electron on one
of the right two atoms and the $\beta$ electron diagonally opposite.
Fig. \ref{fig:orbitals} shows the orbitals for H$_{4}^{2+}$ with
factor 1.27, but the same behavior is present in all H$_{4}^{2+}$
with factors 1 to 2, as the nature of the UHF symmetry breaking is
the same throughout all the systems.

This set of systems offers a collection of very simple molecules where
the Coulson-Fischer point moves along the binding curve and through
the minimum to give a range of different behaviors, all incorrect.
 A molecule such as F$_{2}$ also has an unbound curve with UHF due to symmetry
breaking before the RHF minimum, but is to a large extent fixed by
UB3LYP, which gives a reasonable curve compared to FCI. However a very
key difference in H$_{4}^{2+}$ is that UB3LYP gives a massively too
low total energy due to the fact that the electrons are delocalized
over several centers and the consequent delocalization error dominates,
as illustrated in Fig. 2c and Fig 2d. The same is true for LDA and
GGA functionals.

\section{Conclusions}

Overall, this paper presents simple systems where the UHF total energy
has a very poor behavior compared to FCI upon stretching. For example,
the H$_{4}^{2+}$ rectangular molecule with a ratio of 1.27 gives
a qualitatively incorrect potential energy surface with UHF due to
symmetry breaking at the RHF minimum. This gives rise to a badly predicted
shallow minimum around 2.4 bohr in comparison with the FCI minimum
at 2.1 bohr. This analysis is also very relevant for DFT, where despite
the problems for the stretching of odd electron systems like H$_{2}^{+}$
or He$_{2}^{+}$ \cite{Merkle929216,Ruzsinszky07104102}, there is
a prevailing view that UKS is a good practical solution to the problem
of stretching closed shell molecules \cite{Gunnarsson764274}. This
has been phrased as the symmetry dilemma \cite{Lowdin63496,Perdew954531}.
 In H$_{4}^{2+}$ there is not even the possibility of a dilemma as
UKS calculations are just incorrect due to the delocalization error.
 Therefore, although symmetry breaking is a useful practical tool
in many molecules it should be noted that it can have large failures
in some systems. The observed qualitative breakdown of UHF and the
simple nature of the molecules in which it is found poses a difficult
challenge for all unrestricted methodologies, and hopefully sheds light
on the difficult nature of stretching bonds and related challenges
for electronic structure methods.
\begin{acknowledgments}
We gratefully acknowledge funding from Ramon y Cajal (PMS) and the
Royal Society (AJC). PMS also acknowledges grant FIS2012-37549 from
the Spanish Ministry of Science.
\end{acknowledgments}
\bibliographystyle{aipnum4-1}
%\bibliography{/Users/ajc54/Documents/allnewmodified}

\begin{thebibliography}{28}%
\makeatletter
\providecommand \@ifxundefined [1]{%
 \@ifx{#1\undefined}
}%
\providecommand \@ifnum [1]{%
 \ifnum #1\expandafter \@firstoftwo
 \else \expandafter \@secondoftwo
 \fi
}%
\providecommand \@ifx [1]{%
 \ifx #1\expandafter \@firstoftwo
 \else \expandafter \@secondoftwo
 \fi
}%
\providecommand \natexlab [1]{#1}%
\providecommand \enquote  [1]{``#1''}%
\providecommand \bibnamefont  [1]{#1}%
\providecommand \bibfnamefont [1]{#1}%
\providecommand \citenamefont [1]{#1}%
\providecommand \href@noop [0]{\@secondoftwo}%
\providecommand \href [0]{\begingroup \@sanitize@url \@href}%
\providecommand \@href[1]{\@@startlink{#1}\@@href}%
\providecommand \@@href[1]{\endgroup#1\@@endlink}%
\providecommand \@sanitize@url [0]{\catcode `\\12\catcode `\$12\catcode
  `\&12\catcode `\#12\catcode `\^12\catcode `\_12\catcode `\%12\relax}%
\providecommand \@@startlink[1]{}%
\providecommand \@@endlink[0]{}%
\providecommand \url  [0]{\begingroup\@sanitize@url \@url }%
\providecommand \@url [1]{\endgroup\@href {#1}{\urlprefix }}%
\providecommand \urlprefix  [0]{URL }%
\providecommand \Eprint [0]{\href }%
\providecommand \doibase [0]{http://dx.doi.org/}%
\providecommand \selectlanguage [0]{\@gobble}%
\providecommand \bibinfo  [0]{\@secondoftwo}%
\providecommand \bibfield  [0]{\@secondoftwo}%
\providecommand \translation [1]{[#1]}%
\providecommand \BibitemOpen [0]{}%
\providecommand \bibitemStop [0]{}%
\providecommand \bibitemNoStop [0]{.\EOS\space}%
\providecommand \EOS [0]{\spacefactor3000\relax}%
\providecommand \BibitemShut  [1]{\csname bibitem#1\endcsname}%
\let\auto@bib@innerbib\@empty
%</preamble>
\bibitem [{\citenamefont {Coulson}\ and\ \citenamefont
  {Fischer}(1949)}]{Coulson49386}%
  \BibitemOpen
  \bibfield  {author} {\bibinfo {author} {\bibfnamefont {C.}~\bibnamefont
  {Coulson}}\ and\ \bibinfo {author} {\bibfnamefont {I.}~\bibnamefont
  {Fischer}},\ }\href@noop {} {\bibfield  {journal} {\bibinfo  {journal}
  {Philos. Mag.}\ }\textbf {\bibinfo {volume} {40}},\ \bibinfo {pages} {386}
  (\bibinfo {year} {1949})}\BibitemShut {NoStop}%
\bibitem [{\citenamefont {Savin}(1996)}]{Savin96327}%
  \BibitemOpen
  \bibfield  {author} {\bibinfo {author} {\bibfnamefont {A.}~\bibnamefont
  {Savin}},\ }\enquote {\bibinfo {title} {Recent developments and applications
  of modern density functional theory},}\ \ (\bibinfo  {publisher} {ed J. M.
  Seminario, Elsevier, Amsterdam},\ \bibinfo {year} {1996})\ p.\ \bibinfo
  {pages} {327}\BibitemShut {NoStop}%
\bibitem [{\citenamefont {Cohen}, \citenamefont {Mori-S\'anchez},\ and\
  \citenamefont {Yang}(2008)}]{Cohen08121104}%
  \BibitemOpen
  \bibfield  {author} {\bibinfo {author} {\bibfnamefont {A.~J.}\ \bibnamefont
  {Cohen}}, \bibinfo {author} {\bibfnamefont {P.}~\bibnamefont
  {Mori-S\'anchez}}, \ and\ \bibinfo {author} {\bibfnamefont {W.~T.}\
  \bibnamefont {Yang}},\ }\href@noop {} {\bibfield  {journal} {\bibinfo
  {journal} {J. Chem. Phys.}\ }\textbf {\bibinfo {volume} {129}},\ \bibinfo
  {pages} {121104} (\bibinfo {year} {2008})}\BibitemShut {NoStop}%
\bibitem [{\citenamefont {Mendl}, \citenamefont {Malet},\ and\ \citenamefont
  {Gori-Giorgi}(2014)}]{Mendl14125106}%
  \BibitemOpen
  \bibfield  {author} {\bibinfo {author} {\bibfnamefont {C.~B.}\ \bibnamefont
  {Mendl}}, \bibinfo {author} {\bibfnamefont {F.}~\bibnamefont {Malet}}, \ and\
  \bibinfo {author} {\bibfnamefont {P.}~\bibnamefont {Gori-Giorgi}},\
  }\href@noop {} {\bibfield  {journal} {\bibinfo  {journal} {Phys. Rev. B}\
  }\textbf {\bibinfo {volume} {89}},\ \bibinfo {pages} {125106} (\bibinfo
  {year} {2014})}\BibitemShut {NoStop}%
\bibitem [{\citenamefont {Loos}\ and\ \citenamefont
  {Gill}(2009)}]{Loos09062517}%
  \BibitemOpen
  \bibfield  {author} {\bibinfo {author} {\bibfnamefont {P.-F.}\ \bibnamefont
  {Loos}}\ and\ \bibinfo {author} {\bibfnamefont {P.~M.~W.}\ \bibnamefont
  {Gill}},\ }\href@noop {} {\bibfield  {journal} {\bibinfo  {journal} {Phys.
  Rev. A}\ }\textbf {\bibinfo {volume} {79}},\ \bibinfo {pages} {062517}
  (\bibinfo {year} {2009})}\BibitemShut {NoStop}%
\bibitem [{\citenamefont {Thompson}\ and\ \citenamefont
  {Alavi}(2004)}]{Thompson04201302}%
  \BibitemOpen
  \bibfield  {author} {\bibinfo {author} {\bibfnamefont {D.~C.}\ \bibnamefont
  {Thompson}}\ and\ \bibinfo {author} {\bibfnamefont {A.}~\bibnamefont
  {Alavi}},\ }\href@noop {} {\bibfield  {journal} {\bibinfo  {journal} {Phys.
  Rev. B}\ }\textbf {\bibinfo {volume} {69}},\ \bibinfo {pages} {201302}
  (\bibinfo {year} {2004})}\BibitemShut {NoStop}%
\bibitem [{\citenamefont {Mori-S\'anchez}\ and\ \citenamefont
  {Cohen}(2014)}]{Mori-Sanchez1414378}%
  \BibitemOpen
  \bibfield  {author} {\bibinfo {author} {\bibfnamefont {P.}~\bibnamefont
  {Mori-S\'anchez}}\ and\ \bibinfo {author} {\bibfnamefont {A.~J.}\
  \bibnamefont {Cohen}},\ }\href@noop {} {\bibfield  {journal} {\bibinfo
  {journal} {Phys. Chem. Chem. Phys.}\ }\textbf {\bibinfo {volume} {16}},\
  \bibinfo {pages} {14378} (\bibinfo {year} {2014})}\BibitemShut {NoStop}%
\bibitem [{\citenamefont {Gordon}\ and\ \citenamefont
  {Truhlar}(1987)}]{Gordon871}%
  \BibitemOpen
  \bibfield  {author} {\bibinfo {author} {\bibfnamefont {M.~S.}\ \bibnamefont
  {Gordon}}\ and\ \bibinfo {author} {\bibfnamefont {D.~G.}\ \bibnamefont
  {Truhlar}},\ }\href@noop {} {\bibfield  {journal} {\bibinfo  {journal} {Theo.
  Chim. Acta}\ }\textbf {\bibinfo {volume} {71}},\ \bibinfo {pages} {1}
  (\bibinfo {year} {1987})}\BibitemShut {NoStop}%
\bibitem [{\citenamefont {Purwanto}\ \emph {et~al.}(2008)\citenamefont
  {Purwanto}, \citenamefont {Al-Saidi}, \citenamefont {Krakauer},\ and\
  \citenamefont {Zhang}}]{Purwanto08114309}%
  \BibitemOpen
  \bibfield  {author} {\bibinfo {author} {\bibfnamefont {W.}~\bibnamefont
  {Purwanto}}, \bibinfo {author} {\bibfnamefont {W.~A.}\ \bibnamefont
  {Al-Saidi}}, \bibinfo {author} {\bibfnamefont {H.}~\bibnamefont {Krakauer}},
  \ and\ \bibinfo {author} {\bibfnamefont {S.}~\bibnamefont {Zhang}},\
  }\href@noop {} {\bibfield  {journal} {\bibinfo  {journal} {J. Chem. Phys.}\
  }\textbf {\bibinfo {volume} {128}},\ \bibinfo {pages} {114309} (\bibinfo
  {year} {2008})}\BibitemShut {NoStop}%
\bibitem [{\citenamefont {Nobes}\ \emph {et~al.}(1991)\citenamefont {Nobes},
  \citenamefont {Moncrieff}, \citenamefont {Wong}, \citenamefont {Radom},
  \citenamefont {Gill},\ and\ \citenamefont {Pople}}]{Nobes1991216}%
  \BibitemOpen
  \bibfield  {author} {\bibinfo {author} {\bibfnamefont {R.~H.}\ \bibnamefont
  {Nobes}}, \bibinfo {author} {\bibfnamefont {D.}~\bibnamefont {Moncrieff}},
  \bibinfo {author} {\bibfnamefont {M.~W.}\ \bibnamefont {Wong}}, \bibinfo
  {author} {\bibfnamefont {L.}~\bibnamefont {Radom}}, \bibinfo {author}
  {\bibfnamefont {P.~M.~W.}\ \bibnamefont {Gill}}, \ and\ \bibinfo {author}
  {\bibfnamefont {J.~A.}\ \bibnamefont {Pople}},\ }\href@noop {} {\bibfield
  {journal} {\bibinfo  {journal} {Chem. Phys. Lett.}\ }\textbf {\bibinfo
  {volume} {182}},\ \bibinfo {pages} {216 } (\bibinfo {year}
  {1991})}\BibitemShut {NoStop}%
\bibitem [{\citenamefont {Noodleman}\ and\ \citenamefont
  {Davidson}(1986)}]{Noodleman1986131}%
  \BibitemOpen
  \bibfield  {author} {\bibinfo {author} {\bibfnamefont {L.}~\bibnamefont
  {Noodleman}}\ and\ \bibinfo {author} {\bibfnamefont {E.~R.}\ \bibnamefont
  {Davidson}},\ }\href@noop {} {\bibfield  {journal} {\bibinfo  {journal}
  {Chem. Phys.}\ }\textbf {\bibinfo {volume} {109}},\ \bibinfo {pages} {131 }
  (\bibinfo {year} {1986})}\BibitemShut {NoStop}%
\bibitem [{\citenamefont {Caballol}\ \emph {et~al.}(1997)\citenamefont
  {Caballol}, \citenamefont {Castell}, \citenamefont {Illas}, \citenamefont
  {de~{PR}~Moreira},\ and\ \citenamefont {Malrieu}}]{Caballol977860}%
  \BibitemOpen
  \bibfield  {author} {\bibinfo {author} {\bibfnamefont {R.}~\bibnamefont
  {Caballol}}, \bibinfo {author} {\bibfnamefont {O.}~\bibnamefont {Castell}},
  \bibinfo {author} {\bibfnamefont {F.}~\bibnamefont {Illas}}, \bibinfo
  {author} {\bibfnamefont {I.}~\bibnamefont {de~{PR}~Moreira}}, \ and\ \bibinfo
  {author} {\bibfnamefont {J.~P.}\ \bibnamefont {Malrieu}},\ }\href@noop {}
  {\bibfield  {journal} {\bibinfo  {journal} {J. Phys. Chem. A}\ }\textbf
  {\bibinfo {volume} {101}},\ \bibinfo {pages} {7860} (\bibinfo {year}
  {1997})}\BibitemShut {NoStop}%
\bibitem [{\citenamefont {Jimenez-Hoyos}, \citenamefont {Henderson},\ and\
  \citenamefont {Scuseria}(2011)}]{Jimenez112667}%
  \BibitemOpen
  \bibfield  {author} {\bibinfo {author} {\bibfnamefont {C.~A.}\ \bibnamefont
  {Jimenez-Hoyos}}, \bibinfo {author} {\bibfnamefont {T.~M.}\ \bibnamefont
  {Henderson}}, \ and\ \bibinfo {author} {\bibfnamefont {G.~E.}\ \bibnamefont
  {Scuseria}},\ }\href@noop {} {\bibfield  {journal} {\bibinfo  {journal} {J.
  Chem. Theory Comput.}\ }\textbf {\bibinfo {volume} {7}},\ \bibinfo {pages}
  {2667} (\bibinfo {year} {2011})}\BibitemShut {NoStop}%
\bibitem [{\citenamefont {Knowles}\ and\ \citenamefont
  {Handy}(1988)}]{Knowles883097}%
  \BibitemOpen
  \bibfield  {author} {\bibinfo {author} {\bibfnamefont {P.~J.}\ \bibnamefont
  {Knowles}}\ and\ \bibinfo {author} {\bibfnamefont {N.~C.}\ \bibnamefont
  {Handy}},\ }\href@noop {} {\bibfield  {journal} {\bibinfo  {journal} {J.
  Phys. Chem.}\ }\textbf {\bibinfo {volume} {92}},\ \bibinfo {pages} {3097}
  (\bibinfo {year} {1988})}\BibitemShut {NoStop}%
\bibitem [{\citenamefont {Jarzecki}\ and\ \citenamefont
  {Davidson}(1998)}]{Jarzecki984742}%
  \BibitemOpen
  \bibfield  {author} {\bibinfo {author} {\bibfnamefont {A.~A.}\ \bibnamefont
  {Jarzecki}}\ and\ \bibinfo {author} {\bibfnamefont {E.~R.}\ \bibnamefont
  {Davidson}},\ }\href@noop {} {\bibfield  {journal} {\bibinfo  {journal} {J.
  Phys. Chem. A}\ }\textbf {\bibinfo {volume} {102}},\ \bibinfo {pages} {4742}
  (\bibinfo {year} {1998})}\BibitemShut {NoStop}%
\bibitem [{\citenamefont {Frisch}\ \emph {et~al.}()\citenamefont {Frisch} \emph
  {et~al.}}]{g09}%
  \BibitemOpen
  \bibfield  {author} {\bibinfo {author} {\bibfnamefont {M.~J.}\ \bibnamefont
  {Frisch}} \emph {et~al.},\ }\href@noop {} {\enquote {\bibinfo {title}
  {Gaussian 09 {R}evision {D}.01},}\ }\bibinfo {note} {Gaussian Inc.
  Wallingford CT 2009}\BibitemShut {NoStop}%
\bibitem [{\citenamefont {Dunning}(1989)}]{Dunning891007}%
  \BibitemOpen
  \bibfield  {author} {\bibinfo {author} {\bibfnamefont {T.~H.}\ \bibnamefont
  {Dunning}, \bibfnamefont {{Jr.}}},\ }\href@noop {} {\bibfield  {journal}
  {\bibinfo  {journal} {J. Chem. Phys.}\ }\textbf {\bibinfo {volume} {90}},\
  \bibinfo {pages} {1007} (\bibinfo {year} {1989})}\BibitemShut {NoStop}%
\bibitem [{\citenamefont {Paldus}\ and\ \citenamefont {\ifmmode \check{C}\else
  \v{C}\fi{}i\ifmmode~\check{z}\else \v{z}\fi{}ek}(1970)}]{Paldus702268}%
  \BibitemOpen
  \bibfield  {author} {\bibinfo {author} {\bibfnamefont {J.}~\bibnamefont
  {Paldus}}\ and\ \bibinfo {author} {\bibfnamefont {J.}~\bibnamefont {\ifmmode
  \check{C}\else \v{C}\fi{}i\ifmmode~\check{z}\else \v{z}\fi{}ek}},\
  }\href@noop {} {\bibfield  {journal} {\bibinfo  {journal} {Phys. Rev. A}\
  }\textbf {\bibinfo {volume} {2}},\ \bibinfo {pages} {2268} (\bibinfo {year}
  {1970})}\BibitemShut {NoStop}%
\bibitem [{\citenamefont {Bacskay}(1981)}]{Bacskay81385}%
  \BibitemOpen
  \bibfield  {author} {\bibinfo {author} {\bibfnamefont {G.~B.}\ \bibnamefont
  {Bacskay}},\ }\href@noop {} {\bibfield  {journal} {\bibinfo  {journal} {Chem.
  Phys.}\ }\textbf {\bibinfo {volume} {61}},\ \bibinfo {pages} {385} (\bibinfo
  {year} {1981})}\BibitemShut {NoStop}%
\bibitem [{\citenamefont {Goedecker}\ and\ \citenamefont
  {Umrigar}(1997)}]{Goedecker971765}%
  \BibitemOpen
  \bibfield  {author} {\bibinfo {author} {\bibfnamefont {S.}~\bibnamefont
  {Goedecker}}\ and\ \bibinfo {author} {\bibfnamefont {C.}~\bibnamefont
  {Umrigar}},\ }\href@noop {} {\bibfield  {journal} {\bibinfo  {journal} {Phys.
  Rev. A}\ }\textbf {\bibinfo {volume} {55}},\ \bibinfo {pages} {1765}
  (\bibinfo {year} {1997})}\BibitemShut {NoStop}%
\bibitem [{\citenamefont {Cohen}\ and\ \citenamefont
  {Baerends}(2002)}]{Cohen02409}%
  \BibitemOpen
  \bibfield  {author} {\bibinfo {author} {\bibfnamefont {A.~J.}\ \bibnamefont
  {Cohen}}\ and\ \bibinfo {author} {\bibfnamefont {E.~J.}\ \bibnamefont
  {Baerends}},\ }\href@noop {} {\bibfield  {journal} {\bibinfo  {journal}
  {Chem. Phys. Lett.}\ }\textbf {\bibinfo {volume} {364}},\ \bibinfo {pages}
  {409} (\bibinfo {year} {2002})}\BibitemShut {NoStop}%
\bibitem [{\citenamefont {Knizia}\ and\ \citenamefont {Chan}(2012)}]{FCIcode}%
  \BibitemOpen
  \bibfield  {author} {\bibinfo {author} {\bibfnamefont {G.}~\bibnamefont
  {Knizia}}\ and\ \bibinfo {author} {\bibfnamefont {G.~K.~L.}\ \bibnamefont
  {Chan}},\ }\href@noop {} {\bibfield  {journal} {\bibinfo  {journal} {Phys.
  Rev. Lett.}\ }\textbf {\bibinfo {volume} {109}},\ \bibinfo {pages} {186404}
  (\bibinfo {year} {2012})}\BibitemShut {NoStop}%
\bibitem [{\citenamefont {Mori-S\'anchez}, \citenamefont {Cohen},\ and\
  \citenamefont {Yang}(2008)}]{Mori-Sanchez08146401}%
  \BibitemOpen
  \bibfield  {author} {\bibinfo {author} {\bibfnamefont {P.}~\bibnamefont
  {Mori-S\'anchez}}, \bibinfo {author} {\bibfnamefont {A.~J.}\ \bibnamefont
  {Cohen}}, \ and\ \bibinfo {author} {\bibfnamefont {W.~T.}\ \bibnamefont
  {Yang}},\ }\href@noop {} {\bibfield  {journal} {\bibinfo  {journal} {Phys.
  Rev. Lett.}\ }\textbf {\bibinfo {volume} {100}},\ \bibinfo {pages} {146401}
  (\bibinfo {year} {2008})}\BibitemShut {NoStop}%
\bibitem [{\citenamefont {Merkle}, \citenamefont {Savin},\ and\ \citenamefont
  {Preuss}(1992)}]{Merkle929216}%
  \BibitemOpen
  \bibfield  {author} {\bibinfo {author} {\bibfnamefont {R.}~\bibnamefont
  {Merkle}}, \bibinfo {author} {\bibfnamefont {A.}~\bibnamefont {Savin}}, \
  and\ \bibinfo {author} {\bibfnamefont {H.}~\bibnamefont {Preuss}},\
  }\href@noop {} {\bibfield  {journal} {\bibinfo  {journal} {J. Chem. Phys.}\
  }\textbf {\bibinfo {volume} {97}},\ \bibinfo {pages} {9216} (\bibinfo {year}
  {1992})}\BibitemShut {NoStop}%
\bibitem [{\citenamefont {Ruzsinszky}\ \emph {et~al.}(2007)\citenamefont
  {Ruzsinszky}, \citenamefont {Perdew}, \citenamefont {Csonka}, \citenamefont
  {Vydrov},\ and\ \citenamefont {Scuseria}}]{Ruzsinszky07104102}%
  \BibitemOpen
  \bibfield  {author} {\bibinfo {author} {\bibfnamefont {A.}~\bibnamefont
  {Ruzsinszky}}, \bibinfo {author} {\bibfnamefont {J.~P.}\ \bibnamefont
  {Perdew}}, \bibinfo {author} {\bibfnamefont {G.~I.}\ \bibnamefont {Csonka}},
  \bibinfo {author} {\bibfnamefont {O.~A.}\ \bibnamefont {Vydrov}}, \ and\
  \bibinfo {author} {\bibfnamefont {G.~E.}\ \bibnamefont {Scuseria}},\
  }\href@noop {} {\bibfield  {journal} {\bibinfo  {journal} {J. Chem. Phys.}\
  }\textbf {\bibinfo {volume} {126}},\ \bibinfo {pages} {104102} (\bibinfo
  {year} {2007})}\BibitemShut {NoStop}%
\bibitem [{\citenamefont {Gunnarsson}\ and\ \citenamefont
  {Lundqvist}(1976)}]{Gunnarsson764274}%
  \BibitemOpen
  \bibfield  {author} {\bibinfo {author} {\bibfnamefont {O.}~\bibnamefont
  {Gunnarsson}}\ and\ \bibinfo {author} {\bibfnamefont {B.~I.}\ \bibnamefont
  {Lundqvist}},\ }\href@noop {} {\bibfield  {journal} {\bibinfo  {journal}
  {Phys. Rev. B}\ }\textbf {\bibinfo {volume} {13}},\ \bibinfo {pages} {4274}
  (\bibinfo {year} {1976})}\BibitemShut {NoStop}%
\bibitem [{\citenamefont {L\"owdin}(1963)}]{Lowdin63496}%
  \BibitemOpen
  \bibfield  {author} {\bibinfo {author} {\bibfnamefont {P.~O.}\ \bibnamefont
  {L\"owdin}},\ }\href@noop {} {\bibfield  {journal} {\bibinfo  {journal} {Rev.
  Mod. Phys.}\ }\textbf {\bibinfo {volume} {35}},\ \bibinfo {pages} {496}
  (\bibinfo {year} {1963})}\BibitemShut {NoStop}%
\bibitem [{\citenamefont {Perdew}, \citenamefont {Savin},\ and\ \citenamefont
  {Burke}(1995)}]{Perdew954531}%
  \BibitemOpen
  \bibfield  {author} {\bibinfo {author} {\bibfnamefont {J.~P.}\ \bibnamefont
  {Perdew}}, \bibinfo {author} {\bibfnamefont {A.}~\bibnamefont {Savin}}, \
  and\ \bibinfo {author} {\bibfnamefont {K.}~\bibnamefont {Burke}},\
  }\href@noop {} {\bibfield  {journal} {\bibinfo  {journal} {Phys. Rev. A}\
  }\textbf {\bibinfo {volume} {51}},\ \bibinfo {pages} {4531} (\bibinfo {year}
  {1995})}\BibitemShut {NoStop}%
\end{thebibliography}


\end{document}